\documentstyle[prl,aps,twocolumn]{revtex}
\newcommand{\av}[2]{{\left\langle #1 \right\rangle}_{#2} }

\newcommand{\ep}{\epsilon}
\newcommand{\eq}[1]{ Eq.~(\ref{#1})}

\begin{document}

\title
{\large \bf   Spectral Correlations from the Metal to the Mobility Edge}
\author{D. Braun and G. Montambaux}
\address{Laboratoire de Physique des
Solides,  associ\'e au CNRS \\
Universit\'{e} Paris--Sud \\ 91405 Orsay, France}
\maketitle
\centerline{June 14, 1995}
\vspace*{1.4cm}
\begin{abstract}We have studied numerically the spectral correlations
in a  metallic phase and
at the
metal-insulator transition. We have calculated directly the
two-point
correlation function of the density of states $R(s,s')$.
In the metallic phase, it is well described by the
Random Matrix Theory (RMT). For the first time, we also find numerically the
diffusive corrections for the number variance $\langle\delta n^2(s)\rangle$
predicted by Al'tshuler and Shklovski\u{\i}.
At the transition, at small energy
scales, $R(s-s')$ starts linearly, with a slope larger than in a metal. At
large
separations $|s - s'| \gg 1$, it  is found to
 decrease as a power law $R(s,s') \sim - c /
|s -s'|^{2-\gamma}$ with $c \sim 0.045$ and $\gamma \sim 0.7$, in good
agreement with recent microscopic predictions. At the transition, we have also
calculated the
form factor $\tilde K(t)$, Fourier transform of  $R(s-s')$.  At large $s$,
the number
variance contains two terms
$\langle \delta n^2(s) \rangle= B \langle n \rangle^\gamma +
2 \pi \tilde  K(0)\langle n \rangle$
where $\tilde{K}(0)$ is the limit of the form factor for $t \rightarrow 0$.
\end{abstract}
\pacs{PACS numbers: 1234
 }
%\narrowtext
\section{Introduction}

It is well established that the statistical properties of energy levels in
an isolated metal present universal
features characteristic of the chaotic systems\cite{Efetov83,Altshuler86}. At
low energy
scales, in the so-called ergodic regime, the correlations functions are well
described by the Random Matrix Theory (RMT)\cite{RMT}. In the diffusive
regime, i.e. when the electronic mean free path $l_e$ is smaller than the
typical size $L$ of the system, deviations from RMT occur \cite{Altshuler86}
when
the energy scales become larger than the Thouless energy
 $E_c = \hbar D /
L^2$. This energy is the inverse of the characteristic
time $\tau_D=L^2/D$ for
a particle to diffuse through the sample. $D$ is the diffusion coefficient.
On the other hand, in the localized regime, the correlations between levels
are weaker and in the limit of an infinite system the statistics of energy
levels becomes Poissonnian.

It has been argued that the statistics of energy levels {\it at} the
metal--insulator transition (MIT) is characterized by a third universal
distribution which is a hybrid between the Wigner and the Poisson
distributions. This was clearly shown by a
numerical study of the scaling of the nearest neighbor level spacing
distribution  $P(s)$\cite{Shklovskii93}. Several other
numerical works have confirmed this
idea
\cite{Altshuler88,Hofstetter93a,Hofstetter93b,Hofstetter94a,Sears94,Evangelou94,Zharekeshev94,Varga94}.
Moreover, by using diagrammatic calculations, semiclassical
description  and scaling arguments, it was found that long range
correlations between levels
exhibit a new power--law behavior whose exponent is related to the exponent
$\nu$ of the localization length
 \cite{Kravtsov94a,Aronov94a,Aronov94b,Aronov94c,Kravtsov94c,Kravtsov95}.

There are several quantities which measure the fluctuations of
energy levels. In the RMT,
they depend only on the symmetry of the hamiltonian. If it is invariant
under time reversal symmetry, the fluctuations are described by the
Gaussian Orthogonal Ensemble (GOE) of random matrices (
$\beta = 1$).
When time reversal symmetry is broken, the spectrum
becomes more rigid (Gaussian Unitary Ensemble, GUE, $\beta = 2$).
These quantities are the following:

\begin{itemize}
\item  The number variance $\Sigma^2(E)$
\begin{equation}
\Sigma^2(E)=\langle\delta n^2(E)\rangle = \langle n^2(E) \rangle - \langle n(E)
\rangle^2\,.
\label{variance}
\end{equation}
It measures the fluctuation of the number of levels $n(E)$ in a strip
of width $E$.
 The average $\langle...\rangle$ can be taken either over
different regions of
the spectrum or over several realizations of disorder. In the RMT,
$\Sigma^2(E)$ increases logarithmically with $E$. For $E \gg \Delta$, it
varies as $\Sigma^2(E) \sim (2 / \beta \pi^2)\ln \langle n \rangle$
where $\langle n \rangle = E / \Delta$,
$\Delta$ being the average interlevel spacing.
\item  The distribution
$P(s) = \langle \delta(\ep - \ep_i +\ep_{i+1})\rangle$ of
the spacing $s = \ep/
\Delta $ between consecutive levels. In the RMT, it is well described by the
Wigner--surmise:
$ P(s) \propto s^\beta  \exp(-c_\beta s^2)$,
while, when there is no correlation between levels, it has a Poisson
behavior:
$P(s)= \exp(-s)$.

\item  The two-point correlation function of the Density of States
(DOS)\footnote{This function is usually called $K(\ep)$\cite{Mehta}.
However, we have
chosen to follow here the notations of
 refs.\cite{Kravtsov94a,Aronov94a,Aronov94b,Aronov94c,Kravtsov94c,Kravtsov95}}:
\begin{equation} R(s,s')
= \frac{1}{\rho_0^2}  \av{\rho(\ep)\rho(\ep')}\  -  \
   1\,,\hspace{1cm}
s={\ep \over \Delta} \end{equation}
\noindent where $\rho_0$ is the average DOS.
 Using the definition of the DOS,  $R(s)$ can be simply rewritten as:
\begin{eqnarray}
R(s =\ep/ \Delta)&=&
\langle
\delta(\epsilon-(e_i-e_j))
\rangle - 1     \\
&=&
\delta(s) - 1 + \sum_n P(n,s)   \label{Rdes}
\end{eqnarray}
where $P(n,s)$
is the distribution of distances $s_n$
between levels separated by $n$ other levels\cite{Mehta}. By definition
$P(s)= P(0,s)$.
The
number variance
can  obviously be written in terms of this two-point correlation function:
\begin{eqnarray}
\Sigma^2(E) &=& \int_0^{\langle n \rangle} \int_0^{\langle n \rangle} R(s
 - s') d s d s'\\ &=& 2 \int_0^{\langle n \rangle} ({\langle n
\rangle}-s)R(s) d s\,.\label{sigmaR}
 \end{eqnarray}

\item The form factor $\tilde{K}(t)$, Fourier transform of $R(s)$:
\begin{equation}
\tilde{K}(t) = {1 \over 2 \pi}\int R(s) \exp(i s t) d s \,.      \label{K(t)}
\end{equation}
The interest of this quantity is that is can be directly related to
some characteristics of the classical motion of the diffusive
particles\cite{Argaman93}.
\end{itemize}

Although
the number variance is most frequently used in the literature,
it is  not always the most appropriate quantity to describe the
correlations,
since it is a double integral of the DOS-DOS correlation function
$R(s)$. Thus
the behavior of this quantity at an energy scale $E$ depends on the behavior
of $R(s = \ep /\Delta)$ for {\it all} energies $\ep$
smaller than $E$. That is why in this paper we have chosen to
study $R(s)$ as well as its Fourier transform $\tilde K(t)$ directly.
Although analytical predictions have been given for the asymptotic
behavior of these quantities (large energies or small times) at the MIT
\cite{Kravtsov94a,Aronov94a,Aronov94b,Aronov94c,Kravtsov94c,Kravtsov95},
nothing was known yet about the short range behavior of the
correlations. It is one of the goals of this paper to study
these correlations.
In section 2 we recall
what is their behavior in the metallic regime and compare with numerical
experiments. In section 3, we analyze the two point correlation function at
the MIT. Section 4 is devoted to a discussion of level spacing
distributions, and finally, in section 5, we present our
conclusions.

\section{Spectral correlations in the metallic regime}

Efetov has shown that at energy scales smaller than the Thouless energy
$E_c$, and in the limit where the dimensionless conductance $g = E_c /
\Delta$ is much larger than $1$, i.e. far from a MIT,
 the two-point correlation function $R(s)$ is given by
its expression in the  Gaussian RMT\cite{Efetov83}. This
was confirmed numerically in the framework of the
tight-binding Anderson model with diagonal disorder: the  number variance
and the
distribution $P(s)$ were found to be very well described by the
Wigner-Dyson predictions\cite{Sivan87,Dupuis91}.

We first show  numerical results for the two-point
correlation $R(s)$, fig.\ref{F1}, and its Fourier transform
$\tilde{K}(t)$ in the metallic
regime, fig.\ref{F2}.
 It is seen that the RMT  result is a very good
description of these correlations.
They were obtained by exact diagonalization of a tight binding Anderson
hamiltonian  with on--diagonal disorder via a Lanczos routine. The diagonal
elements are box--distributed around zero with a width $W$, in units of
the transfer integral $t$. We only use
eigenvalues which lie in a central energy strip whose width is chosen such
that it contains approximately half the number of all eigenvalues. Even
though the
average density of states (DOS) is almost constant in this region,
 special care
has to be taken in unfolding the spectrum.
 Then we calculate $R(s)$ defined in eq.\ref{Rdes}.
Thus, $R(\ep)$ is the distribution of differences of eigenvalues in
the energy interval $\Delta E$ used for averaging.  In addition, we average
on different disorder configurations.
The self-correlation of the levels is
 not included in the numerical calculation of $R(s)$.
A $\delta$ function has to be added to the result.
Using rather small systems but
many disorder realizations
($8\times8\times8$
sites and 1000 disorder
realizations for $\ep \leq 5$), we are able to obtain a large amount of
eigenvalues
and correspondingly rather smooth curves for $R(\epsilon)$. As shown in
fig.\ref{F1}b, the numerical precision is even high enough to allow for the
observation of the tiny oscillations in $R(\epsilon)$ in the GUE case
where $R(s) = \delta(s) - \sin^2( \pi s)/(\pi s)^2$.

Let us briefly recall some important features of these correlation
functions.
It is helpful to use a semiclassical argument developped by Argaman {\it et
al.}\cite{Argaman93} in which the function $\tilde{K}(t)$ is directly
related to the
return probability $P(t)$ after a time $t$ for a classical diffusive particle.
More
precisely:
 \begin{equation}
 \tilde{K}(t) =   {2 t P(t) \over 4 \pi^2 \beta} \label{AIS}
 \end{equation}
\noindent The factor $2$
in the numerator arises from interference effects
       between time reversed paths, in the absence of magnetic field or
flux. This relation is valid for classical
times $t$ much smaller than the Heisenberg time $\tau_H = \hbar /
\Delta$ \footnote{For a discussion on the validity of eq.\ref{AIS}
and characteristic times, see ref.\cite{Argaman93}}. For time scales larger
than $\tau_D=\hbar/E_c$, the diffusion is homogeneous so that $P(t)$ is
constant. This
gives a
linear behavior for $\tilde{K}(t)$ as observed in  fig.\ref{F2}. For $t \gg
\tau_H $, $\tilde{K}(t)$
has to saturate. In the energy space, $K(\ep)$ has first a $\delta$ peak at
the origin which describes the self correlation of the levels. Fourier
transform of this peak implies that  $\tilde{K}(t)\to 1/2\pi$ for large $t$.
For small energies, there is a relation between the function $R(s)$ and
the distribution of interlevel spacings $P(s)$:  $R(s)= \delta(s) -1 +
P(s)$ so that
indeed $R(s)$ starts linearly at small $s$ in the absence of a magnetic
field. At
larger energies, $R(s)$ varies like $-1 / (\beta\pi^2 s^2)$. Unless
otherwise specified,
we will now consider the case $\beta=1$, i.e.~no magnetic
field.

 For time scales smaller than $\tau_D$, the diffusion is  non homogeneous
and the classical return probability varies as
\begin{equation}
P(t) = {V \over (4 \pi D t)^{d/2}}
\label{diff}
\end{equation}
so that $\tilde{K}(t)$ scales like $t^{1-d/2}$. For $\epsilon\gg E_c$, its
Fourier transform $R(s)$ is given by:
 \begin{equation}
R(s) \propto  -{1\over s^2} \ \left( {s \over g}\right)^{d/2}  \cos({\pi
d \over 4})\,,  \label{as}
\end{equation}
\noindent and the number variance behaves like $(E/E_c)^{d/2}$ for
$E>E_c \ $  \cite{Altshuler86}.

This "Al'tshuler--Shklovski\u{\i}" power-law behavior had not been seen
numerically
 yet \cite{Hofstetter93b,Hofstetter94a}.  This is because there is a long
cross--over regime around the energy scale $E_c$, and the power
law is expected to  be  visible for very large energy scales $E \gg E_c$.
In the
intermediate regime $E \geq E_c$, the quantization of the diffusion modes
has to be taken into account:
\begin{equation}
P(t) =  \sum_{\bf q} \exp(- D q^2 t)\,,
 \label{diffq}
\end{equation}
Using a large system with
small $E_c$, we have calculated $\Sigma^2(E)$ numerically and found the
onset of the power law.
$\Sigma^2(E)$ starts to be proportional to $(E/E_c)^{3/2}$
only for
energies $E\gtrsim 100E_c$ (see fig.\ref{F3}). If one takes into account
the
fact that for a finite system of size $L^3$ the condition $n\ll L^3$ should
be fullfilled in order to neglect finite
size effects, and if one wants to observe the power law over at least one
decade of energy, the size of the system should be at least of the order
$L^3\sim 10^4$. On the other hand, the discrete sum describes the
numerically calculated $\Sigma^2(E)$ very well, and a fit to this sum
with
$E_c$ as only fitting parameter allows for a  precise determination of
the Thouless energy of the sample  (fig.\ref{F3}).

It turns out that this contribution of the finite-${\bf q}$ diffusion modes
is hardly visible on the two-point correlation function $K(s)$.
For a metal in three dimensions, $K(s)$ varies at
large energies  positively like $+1 / \sqrt{s}$
instead of $-1/s^2$ for RMT, meaning that the levels tend to
attract  each other at
large energies\cite{Jalabert93}. However, these deviations from RMT are very
small,
less than a few percent. We have not been able to see them
on the numerically calculated $K(s)$, see fig.\ref{F1}. For the same
reason, the deviations from the linear behavior of $\tilde K(t)$ for
$t<\tau_d$
are not visible.   Moreover, in the metallic regime that we have studied
($g \gg g_c $), we did not see the finite conductance
deviations  at small
energy scales $s \sim 1$ predicted recently \cite{Kravtsov94b,Andreev95}.
\vspace{.5cm}

\section{The spectral two point correlation function at the
transition} \bigskip

We now turn to the description of the correlations at the MIT. The first
theoretical predictions of spectral correlations at the MIT different from
those in the metallic regime were given by Al'tshuler {\em et
al.}~\cite{Altshuler88}. They predicted a purely linear behavior for the
spectral rigidity,  $\Sigma^2(E) =A \langle n\rangle$ for $\langle
A\rangle\gg 1$, where the
constant $A$ was estimated to be of the order of $0.25$, and thus considerably
different  from the behavior in the  insulating regime where $A$ is
unity. Later, Shklovski\u{\i} {\em et al.}~argued that the level statistics at
the
MIT should be completely universal \cite{Shklovskii93}. They predicted a
level spacing distribution which should be a hybrid between a
Wigner--Dyson type behavior for small $s$, that is $P(s)\propto s$ for $s\to
0$, and a Poisson type for $s\gg 1$: $P(s)\propto\exp(-\alpha s)$. Then
Kravtsov {\em et al.}~brought new dynamics into the field by
showing that the two-point correlation function $R(s)$ has a new long
range power-law behavior: instead of varying as $ -1/( \beta
\pi^2 s^2)$ like
in the Wigner-Dyson regime, it is predicted to obey\cite{Kravtsov94a}:
\begin{equation}
R(s) = - {c_{d\beta} \over s^{2 - \gamma}} \,. \label{tail}
\end{equation}
The non-universal constant $c_{d\beta}$ depends on the dimension and on
the symmetry of the
ensemble, and $\gamma$ is given by
$\gamma=1-1 /(\nu d)$, where $\nu$ is the critical exponent of the
localization length $\xi$. If one uses the
value $\nu\simeq 1.3 - 1.5 $ at the MIT \cite{Kramer93}, the exponent
$\gamma$ is of the order of $\gamma\simeq 0.75$.

This new power-law behavior can be
derived semiclassically from $\tilde{K}(t)$ along the
same lines as in the metallic regime, but now with anomalous diffusion
\cite{Kravtsov94a,Aronov94a,Aronov94b,Aronov94c}. Near
the MIT, the
dimensionless conductance $g = E_c /\Delta$ scales with distance
$\Lambda$ like $g(\Lambda) = g_c [ 1 + (\Lambda/\xi)^{1/\nu}]$, where $\nu$
is the exponent of the localization length $\xi(g)$.
Since
$\Delta$ scales like $\Lambda^{-d}$ and $E_c$ like $\Lambda^{-2}$, this
implies that
the diffusion coefficient has to be size dependent.  For
$\Lambda \ll \xi$ it has to vary as
\begin{equation}
D(\Lambda) \propto  {g_c\over \rho_0} \Lambda^{2-d} \left[ 1+ \left({\Lambda
\over \xi}\right)^{1/\nu}\right]\,.
\end{equation}
\noindent To this length dependent $D(\Lambda)$, one can associate a time
dependent diffusion coefficient $D(t)$,
\begin{equation}
D(t) \propto  \left({g_c\over \rho_0}\right)^{2/d} t^{-1+2/d} \left[ 1+  {2
\over d} \left({g_c t \over \rho_0 \xi^d}\right)^{1/\nu d}\right]\,,
 \end{equation}

\noindent so that according to eqs. \ref{AIS} and \ref{diff}, $\tilde{K}(t)$
is  constant up to a time dependent correction:
\begin{equation}
\tilde{K}(t) \propto {1\over \Delta g_c} \left[ 1- ({L\over
\xi})^{1/\nu}(g_c \Delta t)^{1/\nu d}\right]\,.  \end{equation}
The semiclassical expression (\ref{AIS}) is
known to be equivalent to the two--diffuson diagrammatic  result of
Al'tshuler-Shklovski\u{\i} \cite{Altshuler86}, now with an energy dependent
diffusion coefficient. Aronov {\it et al.}\cite{Aronov94b} argue that at the
transition, there are many more diagramms with the same structure so
that the correct prefactor of the power law is not known, even in sign. This
structure
is supposed to be correct for $t \ll \tau_H$. By Fourier transform, one
has the tail of $R(s)$
\begin{equation}
R(s) \propto - \left({L\over \xi}\right)^{1/\nu}\left({1\over
s}\right)^{1+1/\nu d}\,.
 \end{equation}

At the
transition $\xi \propto L$ so that
one finds the size independent power law
 for the tail of the two-point correlation
function, eq.\ref{tail}.
 % \begin{equation} K(\ep) \propto \sin({\pi \over 2
%\nu d})
 %{1 \over 1+ {1 \over \nu d}}
%\label{tail}
%\end{equation}.
The integration of the power-law tail gives a contribution to
the number variance
$\Sigma^2(E)$ varying as $\langle n\rangle^\gamma$ at
large energies.
The existence of a  linear
term in $\Sigma^2$  was first ruled out
because of a sum rule\cite{Kravtsov94a,Aronov94a}
 --- a statement later discussed and contested
by Aronov and Mirlin \cite{Aronov94c,Kravtsov94c,Kravtsov95}.\\

Finally, Sears and Shore \cite{Sears94} introduced the function
$\Gamma(z)=\Sigma^2/ \langle n\rangle$, which they computed
numerically. With the hypothesis that
the scaling behavior of $\Gamma(z)$ should only depend on the combination of
parameters $z=\langle n\rangle(\xi/L)^d$, they were able to
reproduce the term $\langle n\rangle^\gamma$ in $\Sigma^2$ with the same
$\gamma$ as introduced in \cite{Kravtsov94a}. However, they indeed found an
additional
 linear term. Their numerical results are in  agreement with
$\Gamma (z)=0.30+0.22z^{-2/9}$ for a rather large range of disorder and for
system sizes up to $16\times16\times 16$.

All in all,  a consensus now emerges that for $\langle n \rangle \gg 1$, the
number variance should contain two terms

\begin{equation}
\Sigma^2(E)= A \langle n \rangle + B \langle n \rangle^
\gamma\,.
\end{equation}

Although the power law originates directly from the long range behavior of
the two-point correlation function, the origin of the linear term is more
subtle. It appears as a constant of integration
which is a {\it global} measure of the interaction between eigenvalues
integrated over energy. More precisely, since by definition
(eq.\ref{sigmaR}), $R(s)$ is the second derivative of  $\Sigma^2(s)$,
the coefficient $A$ of the linear term is simply
given by $A = \int_{-\infty}^\infty R(s) d s = 2 \pi \tilde K(0)$. The
absence
of a linear term  in $\Sigma^2(E)$ in the RMT implies that the "sum rule"
$\int_{-\infty}^\infty R(s) d s = 0$ is fulfilled. The
coefficient $A$ contains informations about {\it all
energy scales}, in particular   low energy scales.
However, until now, most studies concentrated on the long range behavior of
$\langle
\delta n^2\rangle$. In the present work, we focus on the {\it direct}
calculation of the two-point correlation function and of the form factor both
for
$s < 1$ and $s \gg 1$.
Using the method described above, we have calculated $R(s)$ for rather
small
system sizes but using many disorder realizations for $W=W_c=16.5$.
In the energy range we considered, finite size effects
can be neglected, which we checked by using different sizes.
In fig.\ref{F6}
we  show  the universal results obtained  from systems with different sizes
in an
energy range $0\le E\le 2$. In this small energy range we used system sizes
up to $16\times 16\times 16$ sites and  up to
1000 disorder realizations. Obviously, the
correlations are weaker than in the GOE and
qualitatively in agreement  with the analytical result of  Kravtsov and
Mirlin \cite{Kravtsov94b} which predicts a correction to the small
energy slope increasing like $1/ g^2$ when $g$ decreases.

The linear variation of $R(s)$ at small $s$ is
consistent with the behavior of $P(s)$ at small $s$ (fig.\ref{F7}).
 More precisely
$R(s) = \delta(s) - 1 + P(s)$. This means that all the $P(n,s)$
for $n \ge 1$ start at least quadratically for
$s \rightarrow 0$, like in the RMT\cite{Mehta}.

For practical purpose , we have fitted $R(s)$ for $s <5$  with
\begin{equation}
R(s)=\delta(s)-\sum_{i=1}^4\,\alpha_i\exp(-s/s_i)
%-\theta(s-s_0)
%{c \over s^{2-\gamma}}
\label{rsum}
\end{equation}
The found coefficients $\alpha_i$ and $s_i$ are shown in table \ref{tab1}.
\begin{table}[h]
\begin{center}
\begin{tabular}{r|l|l|l|l|}
i&1&2&3&4\\ \hline\hline
$\alpha_i$&0.74&0.23&0.02&0.01\\ \hline
$s_i$&0.17&0.51&2.85&3.73\\
\end{tabular}
\caption{Numerical coefficients that describe the short range correlations
(see \protect\eq{rsum}).   \label{tab1}}
\end{center}
\end{table}
It is seen
in fig.\ref{F6} that for $\epsilon\gtrsim 2$, $R(\epsilon)$ is very close to
zero and the correlations become very difficult to find. In order to
detect correlations at larger energies, we have diagonalized a total of
132\,000 systems, the system sizes ranging between $6\times 6\times 6$ and
$8\times 8\times 8$ sites (see table \ref{tab2}).

\begin{table}[h]
\begin{center}
\begin{tabular}{l|c|c|c|}
System&d&c&$2-\gamma$\\ \hline\hline
$6\times 6\times 6$&30\,000&-1.17&0.047\\ \hline
$6\times 7\times 8$&45\,000&-1.12&0.039\\ \hline
$7\times 7\times 7$&20\,000&-1.32&0.046\\ \hline
$7\times 8\times 9$&19\,000&-1.08&0.035\\ \hline
$8\times 8\times 8$&18\,000&-1.14&0.037\\ \hline\hline
average&total: 132\,000&$-1.17\pm 0.08$&$0.041\pm0.005$
\end{tabular}
\caption{Numerical values for the power--law behavior $R(s)\propto
-c/s^{2-\gamma}$ for five different geometries. $d$ is the number of
disorder realizations.   \label{tab2}}
\end{center}
\end{table}
Analyzing the central half
of the spectra (totaling about 23 million levles), we have been
able to extract
the behavior of $R(s)$ for $s \lesssim 25$. In fig.\ref{Rslog} we show the
average correlation function obtained from all systems together with the
best fit of the tail to a power law, given by
\begin{equation}
R(s) \sim  -{0.041 \over s^{1.17}} \label{rs}
\end{equation} in good agreement with the predicted eq.\ref{tail}.
The comparison of the fitted variables for the five different geometries
used allows to estimate the remaining statistical error for the exponent to
$1.17\pm 0.08$ and for the prefactor to $0.041\pm 0.005$ (simple standard
deviations, see table \ref{tab2}). Rather than by statistical errors, the
calculation of $R(s)$ for even higher energies is limited by the precision
of the unfolding of the spectra: We were able to obtain  $\langle
\rho(\epsilon)\rangle$ constant with a precision of about $10^{-4}$. For
$s\simeq 30$, $R(s)$ becomes comparable with this value and thus starts to
fluctuate around zero. This explains the sharp drop in fig.\ref{Rslog} at
these energies and might cause a
further systematic error for $s$ smaller but close to $25$. We estimate the
total error of the exponent to be less than about 20\%.

We now turn to the discussion of the sum rule.
As we have seen above, the coefficient of the linear term in $\Sigma^2$ is
the integral of $R(s)$. It can be written as
 \begin{equation}
A = 1+ 2\int_{0^+}^\infty R(s) d s
 \label{sumrule}
\end{equation}
In the case of the Wigner-Dyson statistics, the weight $\int_{0^+}^\infty
R(s) d s =-0.5$, exactly compensates the weight 1 of the $\delta$ function at
the origin. In the case of a Poisson statistics, $R(s)$ is zero for $s \neq
0$, so that the $\delta$ function is not compensated and there is a linear
term with slope $1$ in $\Sigma^2$. The situation at the MIT is intermediate.

 We see here that
the prefactor is modified by the short range behavior of
$R(s)$. Inserting the numerical values from the fit, we find that
$A(s) = 1- 2
\int_{0^+}^s R(s') d s'$ is about $0.37$ for $s=5$ and  $0.22$
for $s=25$.
Long range correlations reduce $A$ further: {\it assuming}  that one can
extend the obtained power law
for $R(s)$ up to much higher energies, one can
even estimate $A(s)$ for very large energies. It gives
$A \simeq 0.17$ for $s=100$, $A \simeq .09$ for  $s=1000$ and ... $A
\simeq
0$ for  $s \rightarrow \infty$. The sum rule seems to be obeyed for energies
up to infinity. This result constradicts the results of
refs.\cite{Altshuler88} and \cite{Sears94} who predicted
$A(\infty) \simeq 0.25$ and $A(\infty) \simeq 0.30$,  respectively.

We finish this study of the two point correlation function by showing in
fig.\ref{F5}
the numerically calculated struture factor $\tilde{K}(t)$
as a function of time, including energies up to $25\Delta$.
As expected from the above semiclassical argument, it tends to  a finite
value for small $t$ (but still $t$ has to be larger than the collision
time). This finite value is related  to the slope of $\Sigma^2(E)$ for
$E\to\infty$: $\tilde{K}(0)=\lim_{E\to\infty}\partial
\Sigma^2(E)/\partial E$. In our plot, $\tilde{K}(t)$ seems to converge
to the value $\tilde{K}(0)\simeq 0.4$. However, in the numerical evaluation
 of $\tilde{K}(t)$ one
has to use a cut--off in energy. Using $E\simeq 25$, our data
for
$\tilde{K}$ are valid only down to $t\simeq 0.04$, so that we are not able to
calculate properly $\tilde{K}(0)$. For the moment, we can only conclude
that a time dependence of the form

\begin{equation}
\tilde K (t) = \tilde K(0) + b t^{1-\gamma}
 \label{Kdet}
\end{equation}
is compatible with our data  ($b >0$) .

 For the same reason
as in the metallic case (self correlation of levels), $\tilde{K}(t)$ has to
saturate at $\tilde{K}(t)=1/2\pi$ for large $t$.\\

\section{Level spacing distributions at the MIT}

For completeness, we briefly comment on the results concerning $P(s)$.
Shklovski\u{\i} {\it et al.}\cite{Shklovskii93} argued that at large $s$,
it must
follow a "modified" Poisson distribution  of the form $\exp(-\alpha s)$
while
Aronov {\it et al.} predict a tail of the
form $\exp(-A_d\beta
s^{2-\gamma})$. This tail reflects the  behavior of the two-point
correlation function at large $s$.
This is found by describing the thermodynamics of a
one-dimensional gas with a pairwise interaction consistent with the above
two-point correlation function\cite{Aronov94a,Kravtsov94c,Kravtsov95}.
There is no  complete consensus yet on the numerical description of $P(s)$
at the MIT\cite{Hofstetter93b,Zharekeshev94} but  it seems that it can be
globally well fitted by
an expression of the form  $B_1(\gamma) s \exp(-B_2(\gamma)
s^{2-\gamma})$, the constants $B_1$ and $B_2$ being fixed by normalization of
$P(s)$\cite{Evangelou94,Varga94}. Quite recently Kravtsov and Lerner
have constructed a plasma
model with the appropriate power law interaction in order to get the correct
dependence of the two-point correlation function (\ref{tail}). They have
been able to relate the coefficient $B_2$ to the prefactor $c$ of the
power law . They find a value of $c \sim 0.03$ in reasonable
agreement with our result.

 One may wonder if the knowledge of the probability distribution
of distances between non-neighboring levels could add  more useful
information on the long range correlations. For this purpose, we have
 studied the function $P(n,s)$ defined  above.
These functions are known for the RMT regime\cite{Mehta}.
In the Poisson limit, they simply vary as:
 \begin{equation}
P(n,s) = {s^n e^{-s} \over n !}
\end{equation}
It may be much more useful to use these quantities because, for large $s$,
$R(s)$ and $P(s)$ are very small and difficult to calculate.
On the other hand, for large distance $s$, $P(n,s)$ is very different from
the RMT or Poisson cases. Fig.\ref{Pns} shows the successive  $P(n,s)$
at the MIT, and they are compared to the corresponding functions in the Poisson
and RMT
cases. For $s \gg 1$, $P(\langle s \rangle,s)$ characterizes  much
better  the differences between the statistics at the MIT and in the metal
than $P(0,s)$. Unfortunately,  we do not know any prediction for the
behavior of
$P(n,s)$ for large $s,n$ at the MIT.

\section{Conclusions}
We have presented results of the direct numerical calculation of the spectral
two
point correlation function in metals and at the metal--insulator transition.
In the metallic regime the results are in very good agreement with the
RMT predictions. We have been able to find numerically the Al'tshuler and
Shklovski\u{\i} regime
where the diffusion is power-like. At the metal--insulator transition,
 the short
range behavior of the two point correlation
function is clearly modified. The short range part of the correlation is weaker
and is followed
by a power law behavior with an exponent which is consistent with the
picture of anomalous diffusion. At the transition, $R(s)$ is size
independent.
This clearly refutes a prediction\cite{Sears94} that the power--law
contribution should decrease as $L \rightarrow \infty$.
These short range correlations give rise to a non--trivial linear term
in the spectral rigidity $\Sigma^2(E)$ in addition to a power--law term which
 is the signature of the
long range correlations.
By integration of $R(s)$, we conclude that the prefactor of this linear term
may be much smaller than previously anticipated and may even be  zero.

Recently,
different models of interpolating ensembles of matrices have been
 proposed \cite{Pichard94,Moshe94,Canali94,Blecken94}.
Our result puts a constraint on the possible choice of ensembles since they
should produce a
 form factor which varies at small times like eq.\ref{Kdet}.

Acknowledgments: We have benefitted from useful discussions or
correspondance with  E. Hofstetter, V. Kravtsov, I. Lerner, M.L. Mehta and B.
Shklovski\u{\i}. Numerical simulations were performed on the Cray computer of
IDRIS (Institut du D\'eveloppement et des Ressources en Informatique
Scientifique, Orsay).

 \newpage
\begin{figure}[h]

 \caption{ Correlation function $R(s)$ vs $s$ in the metallic
regime ($8\times8\times8$ sites,
$W=4$)  a) with no external flux, b)
  for   $\phi=0.25\phi_0$. The continuous lines are the predictions of
RMT. The energy $E$ is measured in units of $\Delta$, the mean interlevel
spacing. We have omitted the $\delta$ function at the origin}  \label{F1}

\end{figure}

\begin{figure}[h]
  \caption{The Fourier transform $\tilde{K}(t)$ versus  time $t$ in
the
metallic regime  a) with no external flux, b)
  for   $\phi=0.25\phi_0$. The continuous lines are the predictions of
RMT. $t$ is measured in units of $\hbar/\Delta$.}\label{F2}
\end{figure}
\begin{figure}[h]
 \caption{ Number variance $\Sigma^2(E)$ in the metallic regime for a
system with $20\times 20\times 20$ sites and $E_c\simeq 2.5\Delta$ (empty
circles).
This figure shows the cross--over regime between the ergodic (RMT) regime
(dotted line)
and the $E^{3/2}$ regime (full circles). Deviations from the
RMT behavior become visible for $E$ of the order $E_c$. The cross--over
regime is well fitted by the
discrete
sum over modes eq.\protect\ref{diffq} (full line). The onset  of the
asymptotic $E^{3/2}$
behavior is  visible for our finite size system. For larger energies
$\Sigma^2(E)$ decays again due to
finite size effects.} \label{F3}
\end{figure}
\begin{figure}[h]
 \caption{The short range behavior of the correlation function $R(s)$ at the
metal
insulator transition, obtained from systems with sizes  $8\times
8\times 8$ (dots), $14\times 15\times 16$ (stars), and $16\times 16\times 16$
(diamonds). We used up to
1000 disorder realizations. The continuous line is the RMT result (metallic
regime).  We have omitted the $\delta$ function at the origin} \label{F6}
\end{figure}
\begin{figure}[h]
 \caption{The correlation function $R'(s)=R(s)+1-\delta(s)$ together with
$P(s)$. The low energy behavior is identical.} \label{F7}
\end{figure}
\begin{figure}[h]
 \caption{The long range behavior of the correlation function
$-R(s)$ at the metal
insulator transition on a logarithmic plot (full circles), together with the
best fit to a power law (full line). The function was obtained from
132\,000 systems with sizes  between $6\times 6\times 6$ and $8\times
8\times 8$ sites (see table \protect\ref{tab2})} \label{Rslog}
\end{figure}

\begin{figure}[h]
 \caption{Form factor $\tilde{K}(t)$ at the metal--insulator
transition. The continuous line shows the RMT result which fits the metallic
behavior.}\label{F5}
\end{figure}
\begin{figure}[h]
  \caption{$P(n, s)$ in the metal (empty circles), for Poisson (full lines),
and at the
transition (full circles).}\label{Pns}
\end{figure}

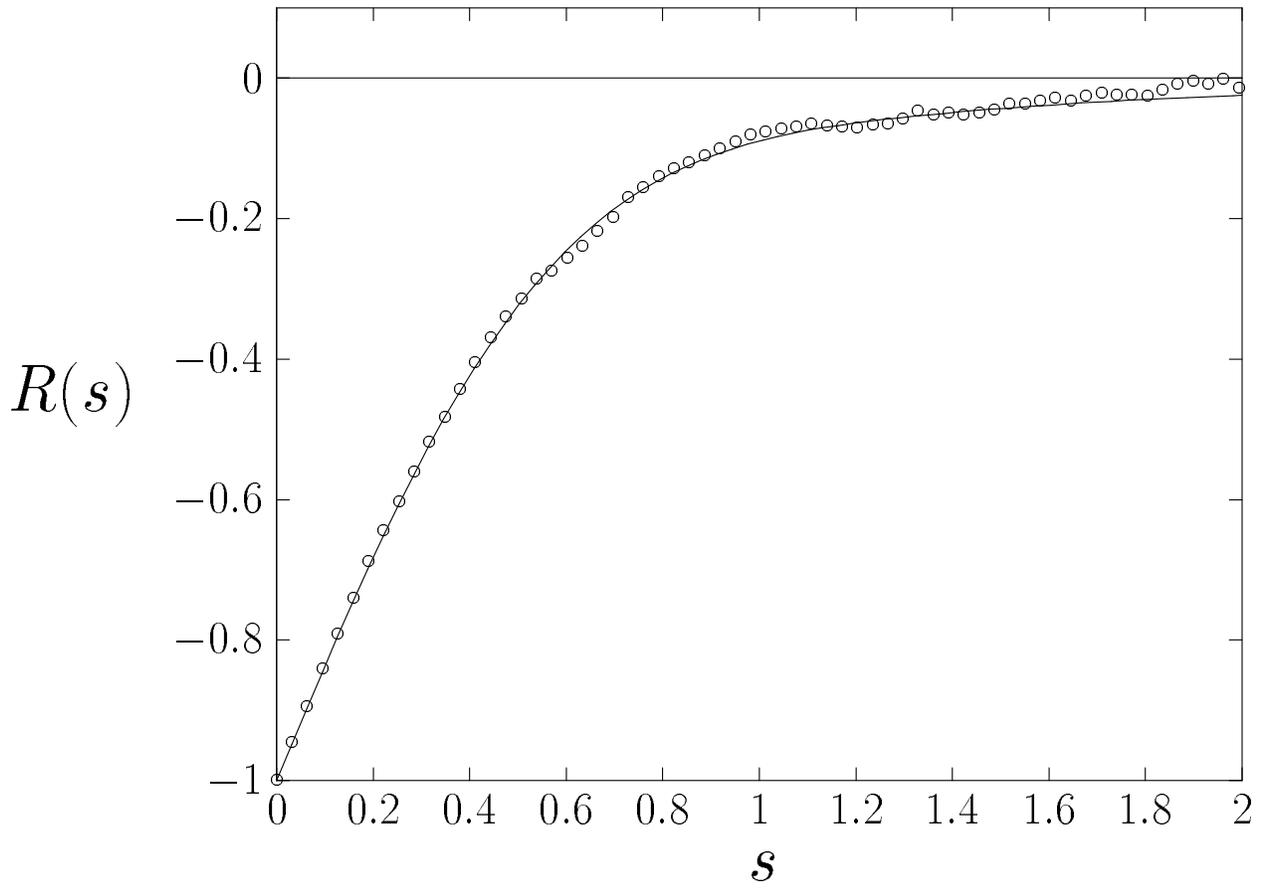
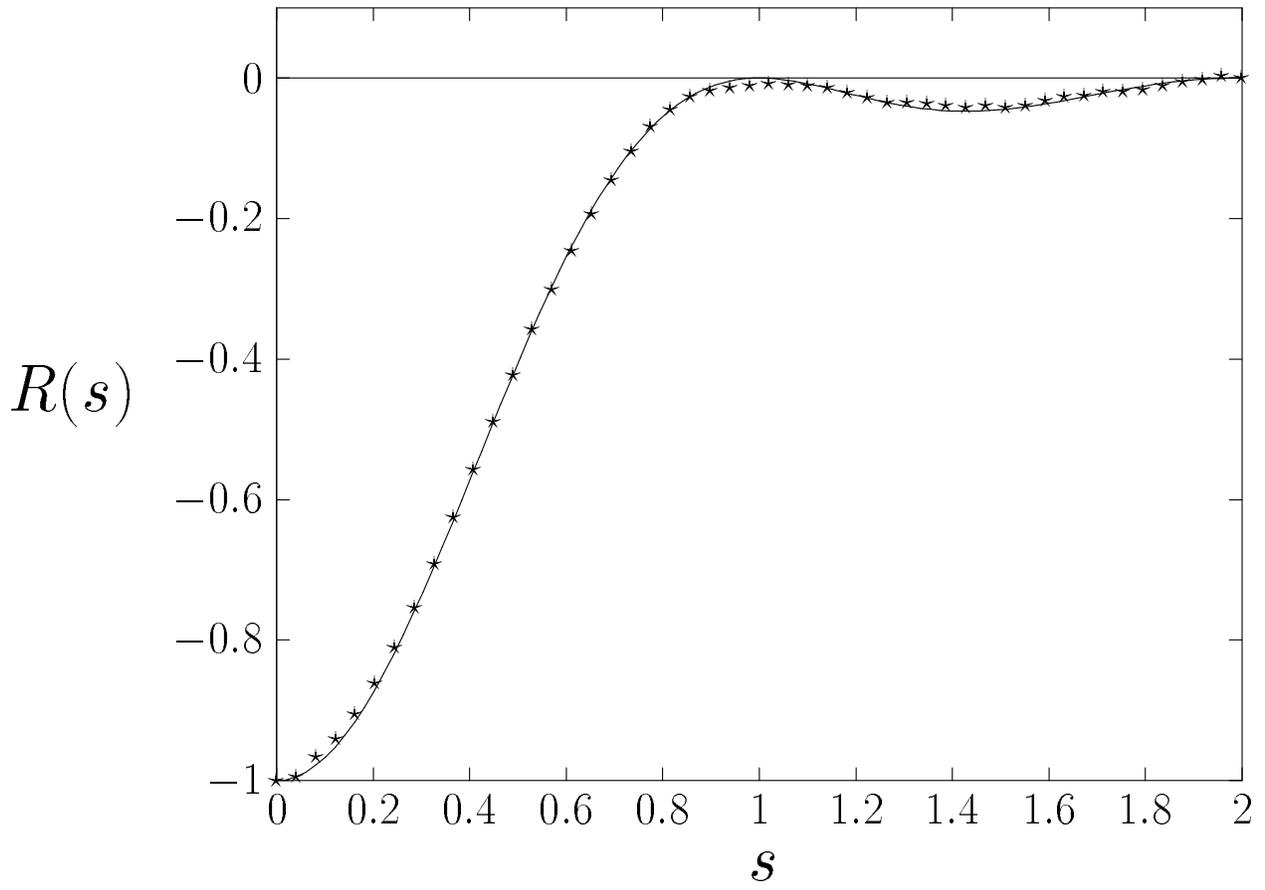
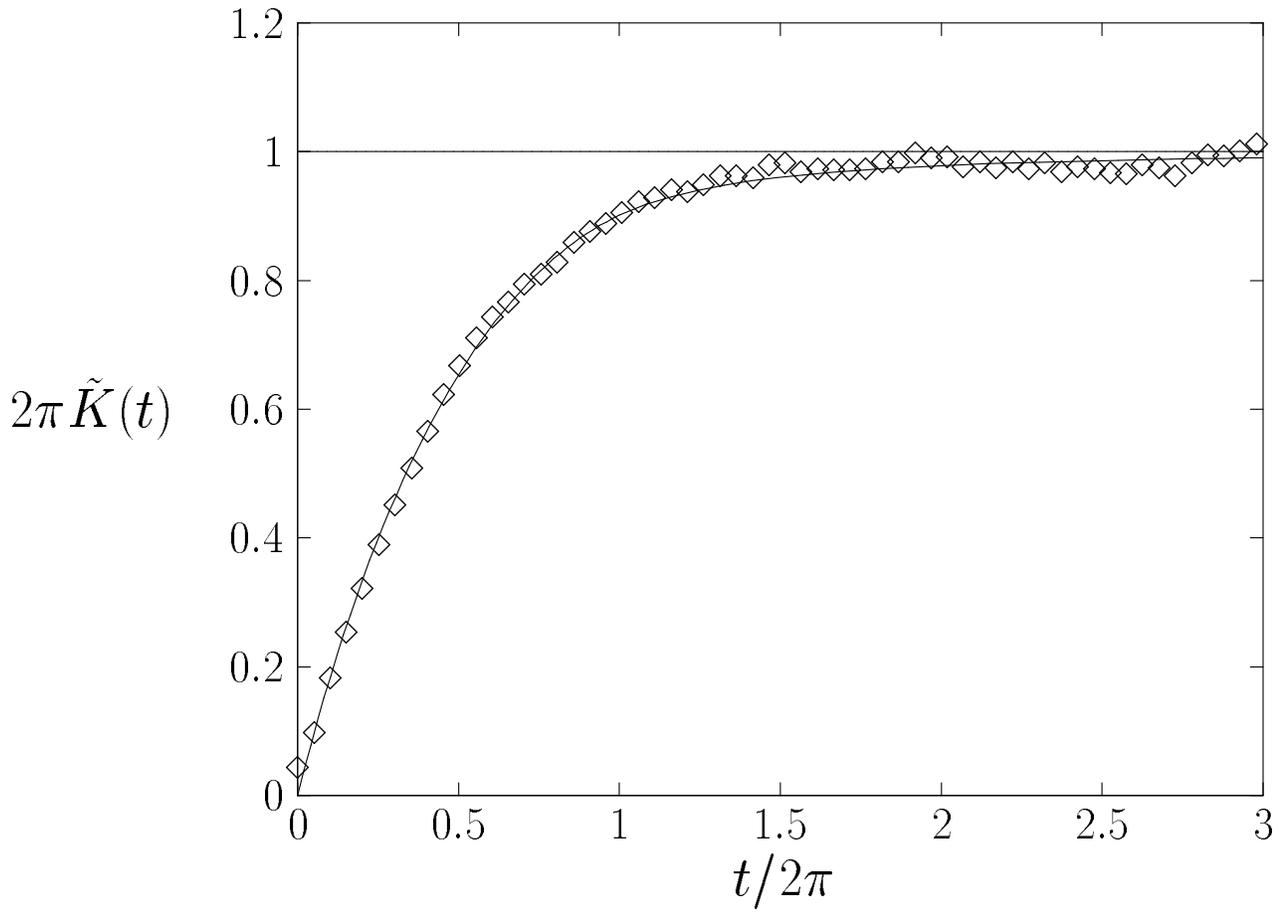
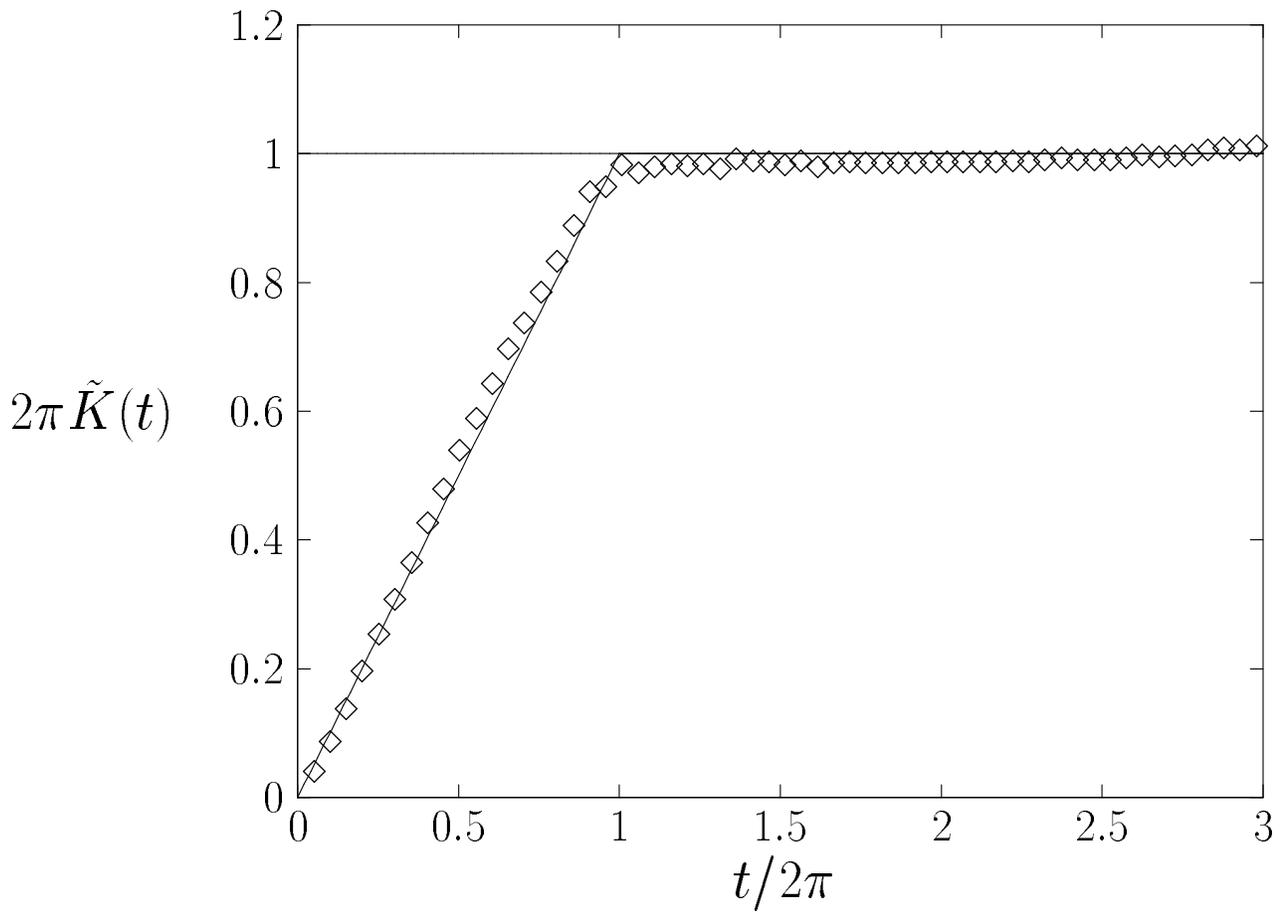
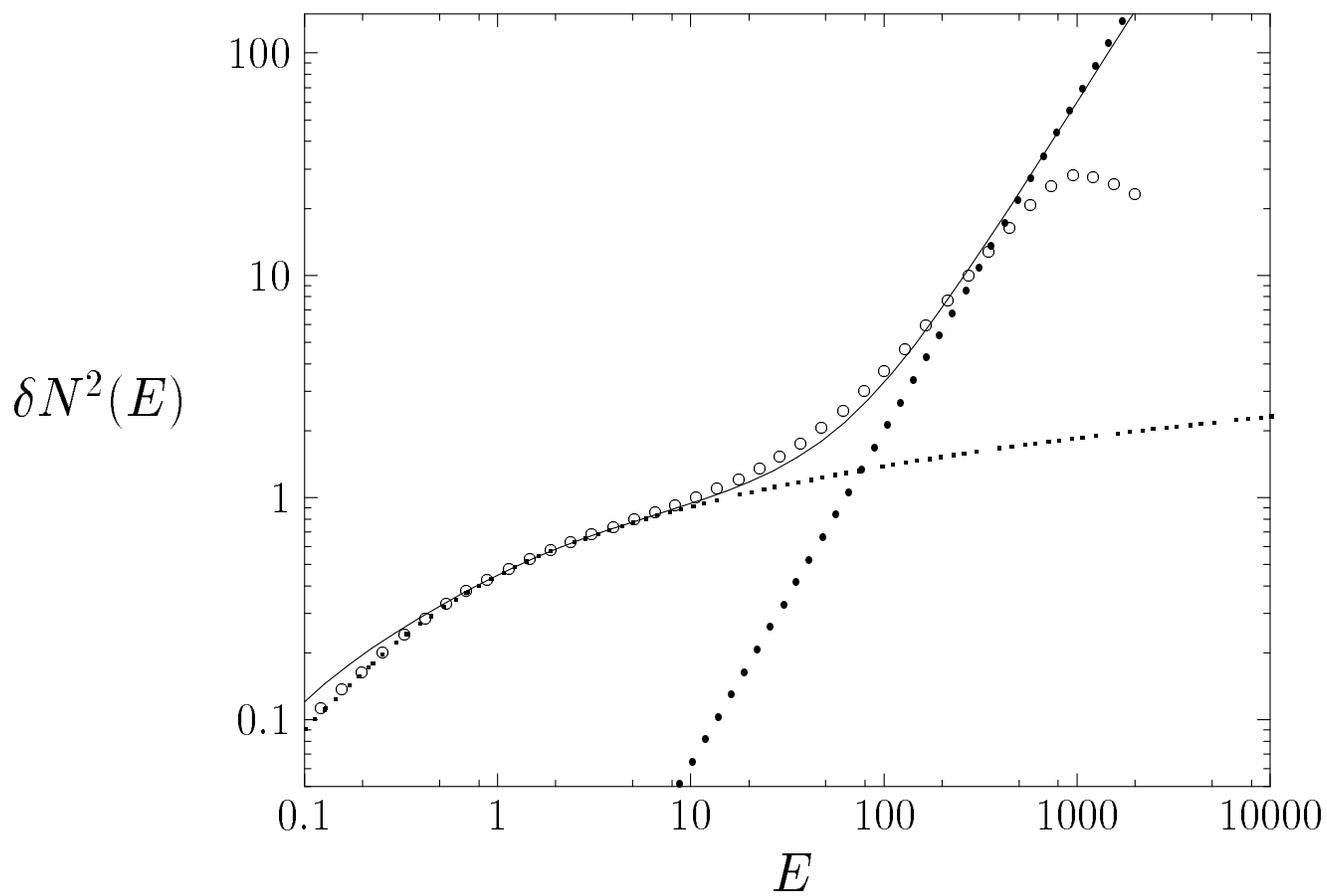
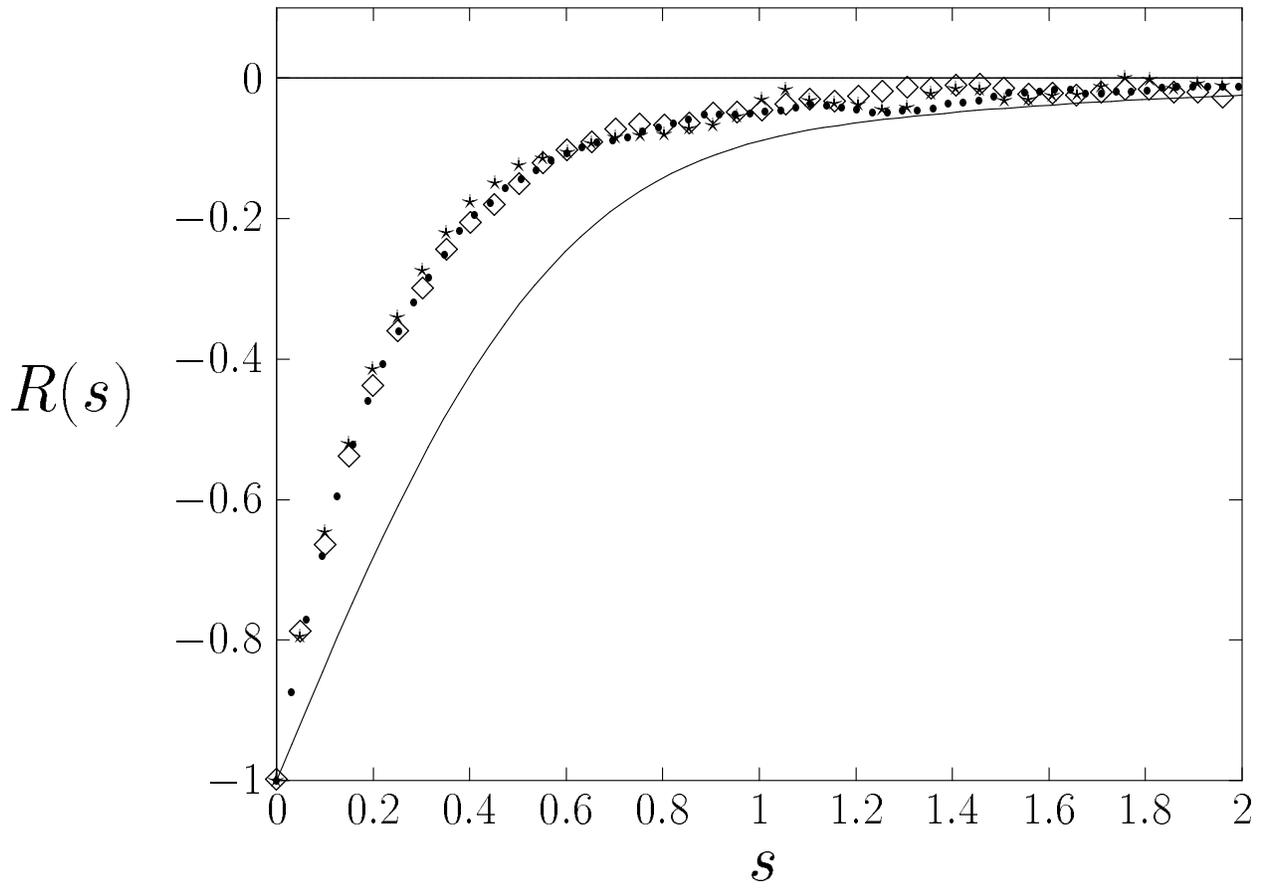
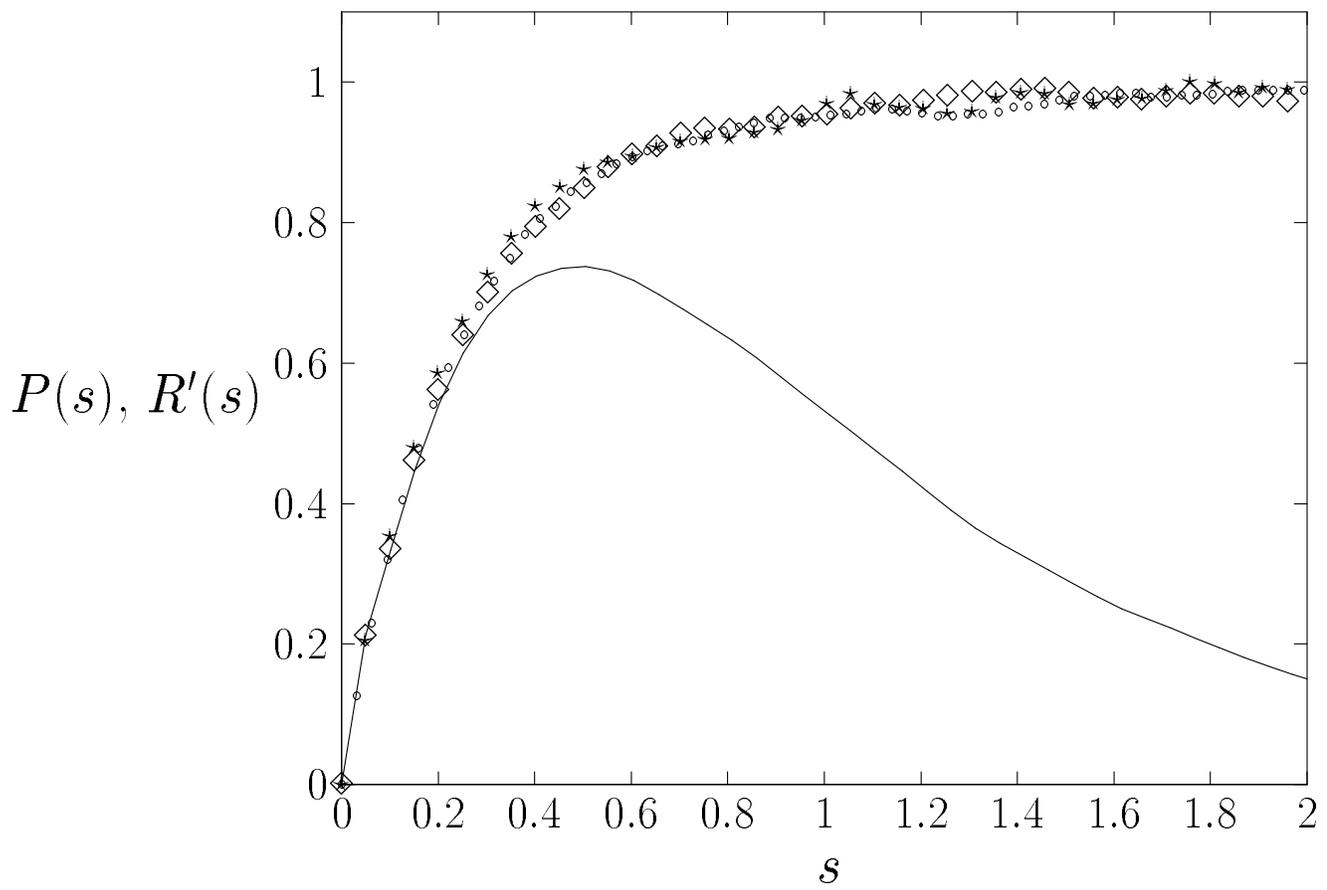
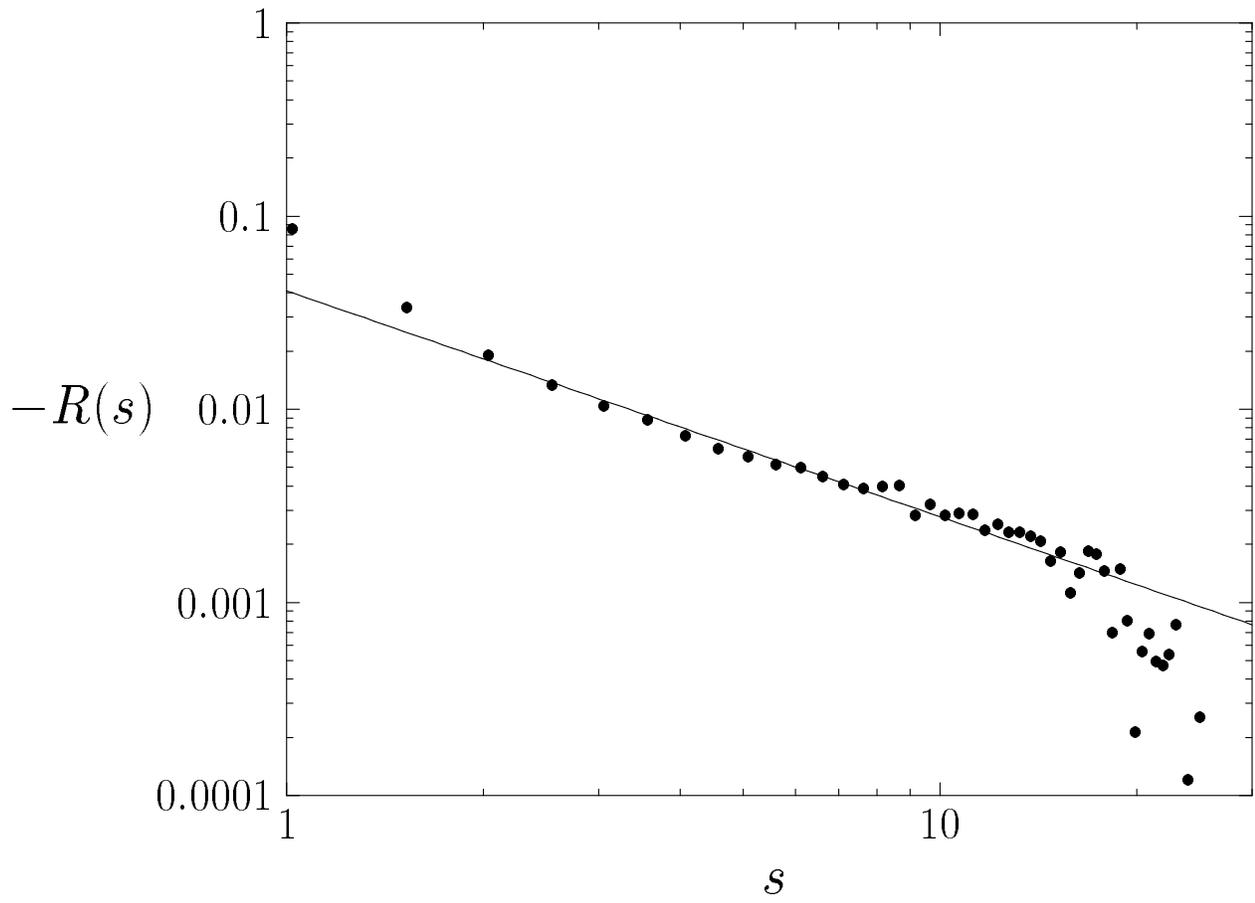
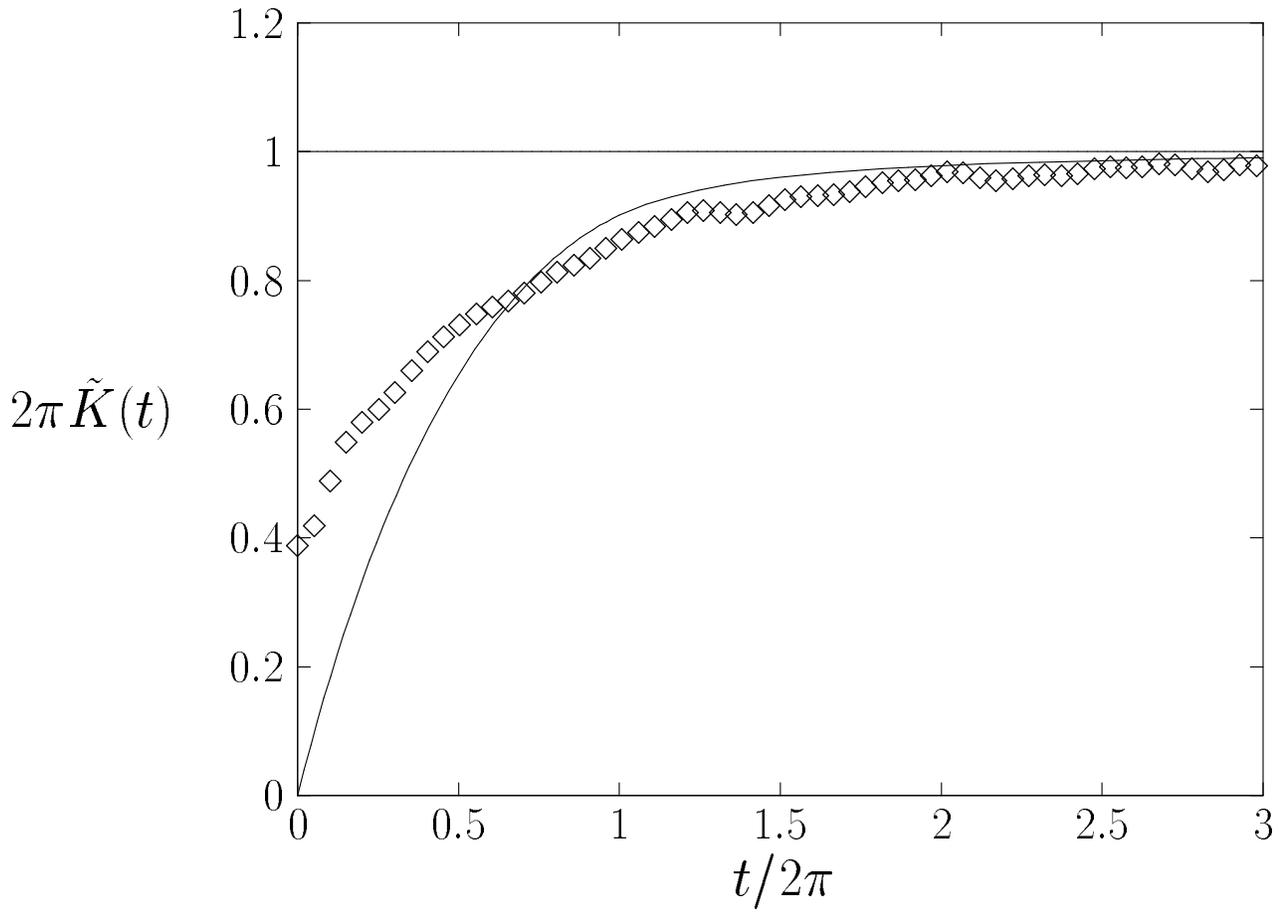
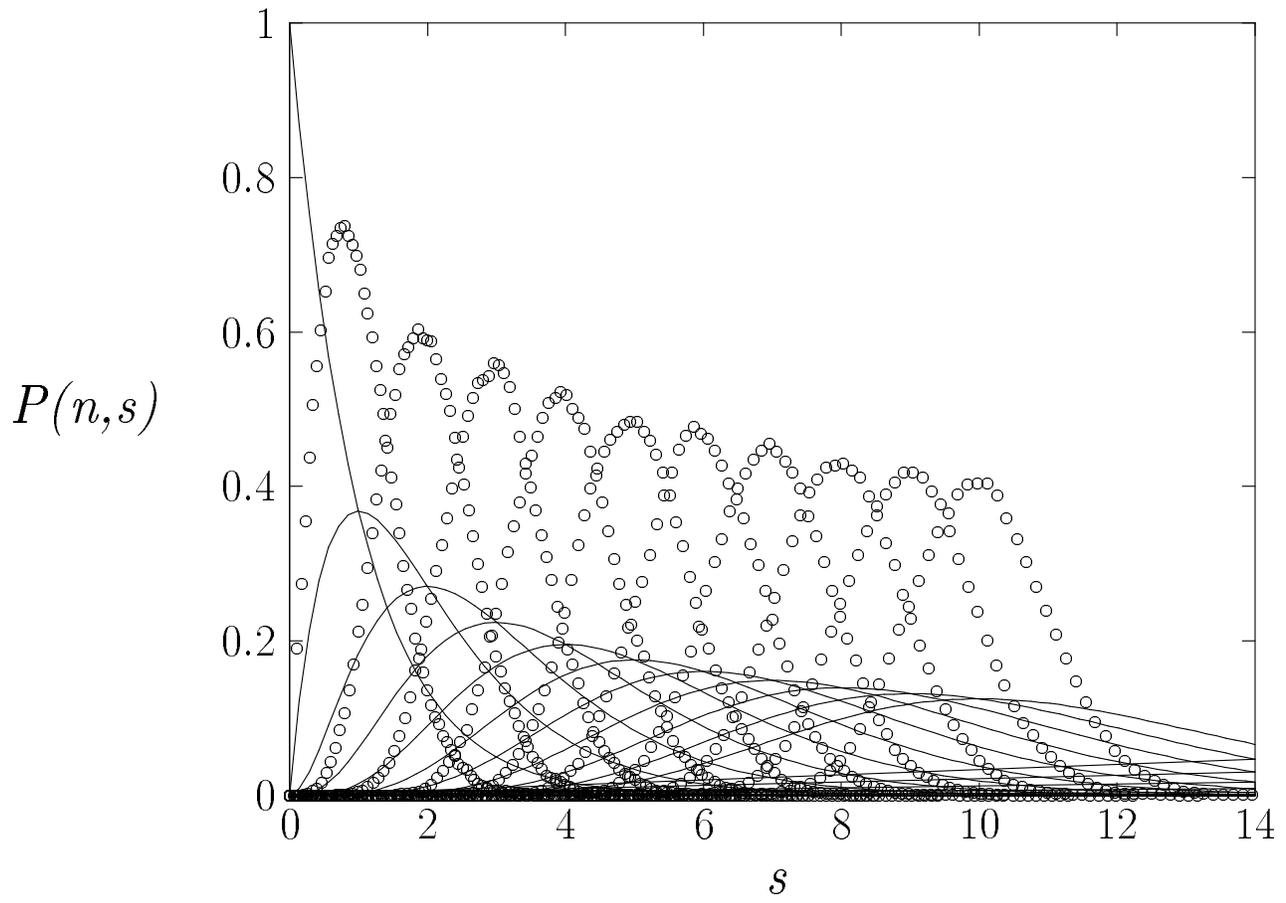
\end{document}